\documentclass{article}

\usepackage{arxiv}

\usepackage[utf8]{inputenc} 
\usepackage[T1]{fontenc}    
\usepackage{hyperref}       
\usepackage{url}            
\usepackage{booktabs}       
\usepackage{amsfonts}       
\usepackage{nicefrac}       
\usepackage{microtype}      
\usepackage{lipsum}
\usepackage{graphicx}
\usepackage{amsmath}
\usepackage{listings}
\usepackage{xcolor}
\usepackage{subfig}
\usepackage{authblk}
\usepackage{multirow}
\graphicspath{ {./images/} }

\definecolor{codegreen}{rgb}{0,0.6,0}
\definecolor{codegray}{rgb}{0.5,0.5,0.5}
\definecolor{codepurple}{rgb}{0.58,0,0.82}
\definecolor{backcolour}{rgb}{0.95,0.95,0.92}
\lstdefinestyle{mystyle}{
    backgroundcolor=\color{backcolour},   
    commentstyle=\color{codegreen},
    keywordstyle=\color{magenta},
    numberstyle=\tiny\color{codegray},
    stringstyle=\color{codepurple},
    basicstyle=\ttfamily\footnotesize,
    breakatwhitespace=false,         
    breaklines=true,                 
    captionpos=b,                    
    keepspaces=true,                 
    numbers=left,                    
    numbersep=5pt,                  
    showspaces=false,                
    showstringspaces=false,
    showtabs=false,                  
    tabsize=2
}
\title{Malceiver: Perceiver with Hierarchical and Multi-modal Features for Android Malware Detection}

\author[ ]{Niall McLaughlin}
\affil[ ]{Centre for Secure Information Technologies  (CSIT)\\ Queen's University Belfast}
\affil[ ]{\textit {n.mclaughlin@qub.ac.uk}}

\begin{document}
\maketitle
\begin{abstract}

We propose the Malceiver, a hierarchical Perceiver model for Android malware detection that makes use of multi-modal features. The primary inputs are the opcode sequence and the requested permissions of a given Android APK file. To reach a malware classification decision the model combines hierarchical features extracted from the opcode sequence together with the requested permissions. The model’s architecture is based on the Perceiver/PerceiverIO which allows for very long opcode sequences to be processed efficiently. Our proposed model can be easily extended to use multi-modal features. We show experimentally that this model outperforms a conventional CNN architecture for opcode sequence based malware detection. We then show that using additional modalities improves performance. Our proposed architecture opens new avenues for the use of Transformer-style networks in malware research.


\end{abstract}


\section{Introduction}

Malware detection for mobile devices, and in particular Android, is a ever growing problem~\cite{McAfee}. In recent years there have been numerous attempts at using deep learning to address this problem. One successful and popular approach has involved static analysis of raw information from executable files, such as raw opcode sequences~\cite{mclaughlin2017deep}, or even the sequence of raw bytes making up an executable file~\cite{raff2018malware}. 

An executable file typically has several levels of hierarchical structure. At the first level we may have sections for the header, resources, and the program code itself~\cite{erdelyi2010reverse}. Then, within each section there is further structure. For instance, the executable code may be structured into classes, methods and low-level code constructs. In this work we focus on Android APK files~\cite{APKFormat}, however analogous hierarchical structure will exist in any executable file format. The information in an executable file can be considered as a sequence containing structured data. The sequence may consist of hundreds of thousands to millions of elements. To perform malware analysis on this kind of data requires a system capable of processing long sequences and uncovering long-range structured patterns. 

The standard deep-learning approaches for sequence processing are recurrent networks e.g., LSTM~\cite{hochreiter1997long}, or Transformers~\cite{vaswani2017attention}. When applied to very long sequences, such as long opcode sequences, the standard approaches may have impractically long training and inference times. This has led to the use of efficient models such as 1D convolutional neural networks (CNNs) for opcode-sequence based malware detection~\cite{mclaughlin2017deep,raff2018malware}. While 1D CNNs have had success in this area, they remain fundamentally local models that operate in a feed-forward, bottom-up, manner. CNNs detect local patterns in the input which they aggregate to detect higher-level or large-scale patterns. Finally, they use a fixed decision procedure to classify samples based on the local patterns detected. There is no mechanism for combining top-down and bottom-up reasoning, which may be needed to handle complex cases where the interpretation of evidence depends on context.

\begin{figure}[t!]
\centering
\includegraphics[width=\textwidth,trim={0 0.75cm 0 2cm},clip]{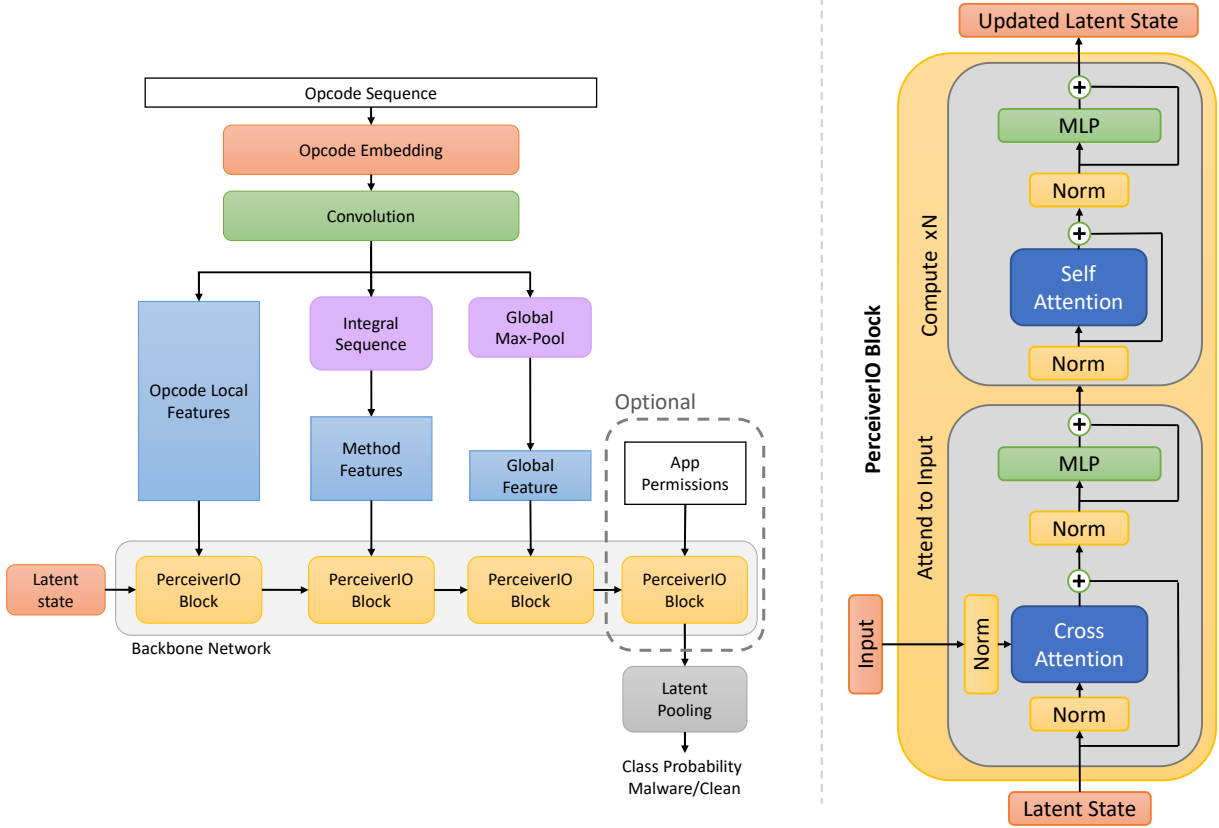}
\caption{Left - Overview of Malceiver Network architecture, showing optional additional PerceiverIO block for multi-modal features. Right - Detailed view of a single PerceiverIO block (Note the block is rotated 90 degrees relative to the figure on the left). The Malceiver processes an opcode sequence using features from the opcode, method and global levels. It uses cross-attention within each PerceiverIO block to combine information across the feature hierarchy. The architecture can be extended via additional PerceiverIO blocks to accommodate multiple input modalities e.g., requested permissions.\label{fig:network_arch}}
\end{figure}

To address the shortcomings of existing methods, we propose an efficient multi-modal and hierarchical transformer-based model for Android malware detection using static analysis (See Fig.~\ref{fig:network_arch}). Our proposed architecture is based on the Perceiver/PerceiverIO~\cite{jaegle2021perceiverIO,jaegle2021perceiver} architectures. The Perceiver makes it practical to apply self-attention to very long sequences in an efficient manner. This allows long-range dependencies to be learned and allows reasoning over local and global information. In our model, we take into consideration the hierarchical structure of programs and we show how to efficiently extract features representing each method using Integral Sequences. We also show how our model can be extended to use multi-modal features by combining opcodes-sequences with requested permissions, which improves performance. Finally, our proposed architecture is more flexible than existing approaches. It has the potential to be extended to include many additional modalities and the network can grow to take advantage of increasingly large and diverse datasets.

\section{Related Work}

\paragraph{Malware Detection} The enormous growth malware in recent years~\cite{McAfee}, and the fact that manual analysis of malicious applications does not scale well, has spurred research into automated methods for malware analysis and detection. As a result, there is now a significant body of work in applying machine-learning / deep-learning to the malware detection problem~\cite{liu2021deep}.

At first, traditional machine-learning methods were explored. Such methods typically require complex pipelines consisting of hand-designed feature extraction, feature-selection, dimensionality-reduction and classification steps. Given pre-processed features, classifiers such as, KNN~\cite{saracino2016madam}, SVMs~\cite{arp2014drebin} or decision trees~\cite{aafer2013droidapiminer} could then be applied. While such traditional approaches can be successful, they require significant engineering effort, and ongoing support to keep pace with the ever changing malware landscape e.g., by designing new feature extraction methods.

Following the success of deep learning in numerous fields, such as visual object recognition~\cite{Krizhevsky2012ImageNetCW}, speech recognition~\cite{benesty2008automatic}, and natural language processing~\cite{vaswani2017attention}, neural network based methods have become the dominant approach to malware classification in recent years~\cite{liu2021deep}. In ideal circumstances, deep-learning removes the need for feature-engineering and for complex software pipelines. Instead, feature-extraction and classification can be performed by a single neural network trained end-to-end on a large dataset of raw data. This approach has the potential to improve malware detection performance by jointly optimising the entire pipeline at once~\cite{liu2021deep}.

Some early steps in applying deep-learning to malware detection combined manual feature engineering with neural network classifiers e.g., Droid-Sec~\cite{yuan2014droid}. Later approaches started to train feature-extraction and classification in a more end-to-end manner such as DroidDetector~\cite{yuan2016droiddetector} and MalDozer~\cite{karbab2018maldozer}. 

A popular malware detection approach, followed in this paper, has been static-analysis of opcode-sequences. The opcode sequences are typically analysed using convolutional neural networks (CNNs). This follows from work in NLP showing that CNNs can perform sequence processing tasks traditionally requiring specialised hand-designed pipelines~\cite{chen2015convolutional}. In one strand of research in this area, the opcode sequences are reshaped to 2D images~\cite{hsien2018r2,khan2019analysis,kumar2018malicious} for classification by CNNs originally designed for visual object classification e.g., Alexnet or ResNet~\cite{Krizhevsky2012ImageNetCW, he2016deep}. However, such approaches may learn spurious correlations between unrelated parts of programs due to the reshaping operation applied to the opcode sequence. An alternative approach, followed here, uses sequence-based CNNs to process the opcode sequence directly~\cite{mclaughlin2017deep,raff2018malware}. This later approach matches the network design to the underlying sequential nature of the opcode-sequence data~\cite{mclaughlin2017deep}. Recent work has shown how opcode-sequence data can be complemented with additional modalities for improved classification performance~\cite{millar2021multi}, and shown how the robustness of these methods to obfuscation can be improved~\cite{millar2020dandroid}.

\paragraph{Efficient Transformer Models}

The transformer model was originally proposed for NLP tasks~\cite{vaswani2017attention}. The popularity of this architecture has grown in recent years, with a variety of transformer inspired models being applied to other tasks such as visual object classification~\cite{dosovitskiy2020image}. The core of the transformer architecture is the self-attention mechanism, which allows long-range non-local information passing~\cite{vaswani2017attention}. The disadvantage of self-attention is its high computational complexity. This is caused by the need to compare all elements of the input sequence with all others, leading to $O(n^2)$ complexity in computation and time. This restricts the transformer to be used with relatively short input sequences. Standard transformer models are limited to processing sequences with lengths of order $10^3$ elements~\cite{tay2020efficient}. To overcome this limitation there has been recent interest in creating more efficient transformer models.

Many approaches to creating efficient transformer models have focused on removing the need for self-attention to compare every element of the input with all others. For example, sparse attention~\cite{correia2019adaptively}, random attention~\cite{weissenborn2019scaling} and learned attention masks~\cite{kitaev2020reformer}. There have also been approaches based on learning low-rank kernels to approximate the self-attention operation~\cite{peng2021random}. Despite increases in efficiency compared with the original transformer, most of the above approaches still cannot be applied to extremely long input sequences e.g., sequences with over $10^5$ or $10^6$ elements, which is needed for processing full-length opcode-sequence.

A recent pair of promising approaches, the Perceiver~\cite{jaegle2021perceiver} and PerceiverIO~\cite{jaegle2021perceiverIO}, change the attention mechanism to use cross-attention rather than self-attention. By doing so they can accommodate extremely long input sequences. The PerceiverIO has been shown to be effective even with input sequences with over 2 million raw elements~\cite{jaegle2021perceiverIO}. These architectures are suitable for applying the transformer to long opcode sequences for malware detection. The Perceiver architecture has also been recently extended with the Hierarchical Perceiver, which aims to further improve efficiency~\cite{carreira2022hierarchical}. This differs from the hierarchical Perceiver model proposed here as we split the input in a way that is specifically adapted to the needs of the malware domain.

We propose a new network architecture, the Malceiver, that adapts the Perceiver/PerceiverIO architecture to Android malware detection. Our model takes the hierarchical structure of programs into consideration by combining information from multiple levels of the program feature hierarchy, thus improving performance compared with a baseline model using raw opcode sequences only. We  show how this architecture can be extended to use multi-modal feature fusion for improved malware classification performance. Finally, we believe that our model opens new research avenues, allowing powerful Transformer-style networks to be applied to malware classification. The proposed model is highly flexible and can grow to accommodate ever larger malware datasets and many different multi-modal feature types. 


\section{Method}

Our model is based on the Perceiver/PerceiverIO~\cite{jaegle2021perceiver,jaegle2021perceiverIO} family of architectures. We first show how our proposed network can process an extremely long opcode sequence, which may contain over $10^5$ elements. We later show how this model can be extended to use multi-modal features (see Section~\ref{section:multimodal}) that need not be sequential themselves.


\subsection{Input}

The network takes as input a sequence of opcodes, $X=\{x_1,\dots,x_n\}$. This sequence is obtained by first disassembling an Android APK file. Given the disassembly, the operands are discarded leaving only the sequence of opcodes. The opcode sequences from all methods and all classes are then concatenated to produce the complete opcode sequence for the given program. In some cases the complete sequence will contain over $10^6$ opcodes.

We consider that programs are organised hierarchically: into packages, classes, methods and code constructs. Disassembly of an Android executable, via tools such as APKtool~\cite{apktool}, provides access to the program structure. Given the complete opcode sequence, $X$, and metadata from the disassembly, we record the starting and ending indexes of each method within the opcode sequence. This produces a structure $\mathcal{F} = ((s_1,e_1),(s_2,e_2),\dots,(s_f,e_f),\dots,(s_F,e_F))$, where $(s_f,e_f)$ are the starting and ending indexes of method $f$. The indexes point to the starting and ending opcodes of each method within the overall opcode sequence $X$. We will later use this information to extract features from every method, known as \emph{method features} (See Section~\ref{sec:method_features}).

\subsection{Hierarchical Feature Extraction}

In this section we show how the raw opcode sequence can be processed to produce a feature hierarchy. This allows the program to be analysed at three levels: at level of raw opcodes, at the level of individual methods, and at the global level.

\subsubsection{Opcode Level Features}

Building on existing work in both efficient vision-based Transformer models~\cite{dosovitskiy2020image} and CNN models for opcode-sequence based malware detection~\cite{mclaughlin2017deep}, we first process the raw opcode sequence using an opcode embedding layer and 1D CNN network. These layers have several purposes. Firstly, the embedding layer allows the network to learn its own vector embedding representation for individual opcodes, which encodes their semantics. Secondly, the CNN layer helps to reduce the amount of training data needed by the transformer backbone~\cite{hassani2021escaping}. It does this by introducing an inductive bias that opcodes are related to their neighbours within each local region of the sequence. We note that this does not prevent the overall network from learning long-range or global relationships. Such non-local relationships can be learned by the Perceiver block layers later in the network.

The input opcode sequence, $X$, is first processed by the opcode embedding layer to produce a matrix, $P\in\mathcal{R}^{n\times e}$, where, $n$ is the sequence length and $e$ is the opcode embedding dimensionality. The embedding layer maps each opcode to a corresponding embedding vector. It allows the network to learn a semantic representation for each opcode. Next the matrix $P$ is passed through by a 1D convolutional layer with a ReLU activation function. This produces a matrix $C\in\mathcal{R}^{n\times c}$ where $c$ is the number of convolutional filters and $n$ corresponds to the sequence length. Each convolutional filter is of length $c_l$ and width $e$ i.e., the same width as the input embedding matrix $P$. Note that the overall sequence length, $n$, remains unchanged after the embedding and CNN layers. There is a one-to-one correspondence between rows of $C$ and the original opcode sequence. We interpret each row, $c_i \in C$, as an opcode neighbourhood embedding vector that contains semantic information about all the surrounding opcodes within its local neighbourhood. The size of the local neighbourhood depends on the length $c_l$ of the 1D convolutional filters used.


\subsubsection{Extracting Method-Level Features using Integral Sequences}
\label{sec:method_features}

Given the matrix $C$ output by the convolutional layer, we now show how features can be extracted to summarise the high-level functionality of each method. If all methods within the program were of equal length, we could simply apply a standard pooling operation with a fixed window-length and stride. However, in reality, methods differ greatly in their lengths, meaning we require a different approach. We require an operation that is efficient enough to be used with extremely long opcode sequences without slowing down training.

We build on the integral-images approach, applied in the Viola Jones face detector~\cite{viola2001robust}. For face detection this method efficiently computes the sum of all pixel values within a specified region of an image. Here we propose to use the Integral Image approach to efficiently find the summation of all opcode embedding vectors within a method to summarise the method's behaviour. We note that in natural language processing (NLP), the summation of word embedding vectors can be used to compute sentence embeddings for summarising the overall meaning of sentences~\cite{mikolov2013distributed,arora2017simple}.

The output from the convolution layer is the matrix $C\in\mathcal{R}^{n\times c}$, where $c$ is the embedding dimension and $n$ is the program length dimension. We first compute the cumulative sum along the program length dimension. This gives the matrix $\hat{C}\in\mathcal{R}^{n\times c}$. Then, to compute a summary feature for every method we refer to $\mathcal{F}$, which holds the starting and ending indexes of every method. For a given method $f$, where $(s_f,e_f) \in \mathcal{F}$ are its starting and ending indexes, we compute the feature $\hat{f} = (\hat{C}_{e_f} - \hat{C}_{s_f}) / (e_f - s_f)$, where $\hat{C}_j$ denotes row $j$ of matrix $C$. Each feature $\hat{f}\in\mathcal{R}^c$ now summarises the behaviour of a particular method. Method features are normalised by dividing by the method length. Finally, the features from all methods are concatenated to produce a matrix $M\in\mathcal{R}^{F \times c}$, where $F$ is the total number of methods in the program.

\subsubsection{Global Feature Extraction}

To give our network the ability to combine local with global evidence, we extract a single feature summarising overall global program behaviour. Given matrix, $C$, output from the convolutional layer, we perform global max-pooling along the program length dimension. This produces a single feature vector $g \in \mathcal{R}^{1\times c}$ representing global program behaviour.

\subsection{PerceiverIO Network}

We have now shown how hierarchical representations of the opcode sequence can be extracted at the opcode-level, method-level and global-level. These representations consist of sequences of embedding features with different sequence lengths but identical embedding dimensionality. The opcode-level features may have a sequence length of over $10^6$ i.e., one feature for every opcode. The method-level feature sequence may contain around $10^5$ features. Finally, a single feature vector summarises the global program behaviour. We will now show how the information from all levels of the opcode sequence feature hierarchy can be combined.

The standard transformer architecture~\cite{vaswani2017attention} uses a self-attention mechanism where every element of an input sequence is compared with every other element. This allows for long-range information passing and reasoning over the whole sequence length. However, self-attention has $O(n^2)$ computational and memory complexity with respect to sequence length. Therefore processing long sequences i.e., ones with more than a few thousand elements, is impractical given the standard self-attention mechanism~\cite{jaegle2021perceiver}.

To process opcode sequences, which may have over $10^6$ elements, we require a different approach. We therefore build on the Perceiver/PerceiverIO network architecture~\cite{jaegle2021perceiver,jaegle2021perceiverIO}. In contrast with self-attention, the PerceiverIO uses cross-attention with a learned latent state matrix. It first reads relevant information from the input sequence using cross-attention between the input sequence and the PerceiverIO's initial latent state matrix. This produces an updated latent state matrix containing relevant information. The computational complexity of this cross-attention operation is $O(nm)$ where $n$ is the length of the input sequence, $m$, is the length of the PerceiverIO's latent state matrix, and $m<<n$. Next the Perceiver performs self-attention on the updated latent state matrix, which contains a summary of the input. Given the small size of the latent state matrix the self-attention operation is more efficient than operating directly on the input.

In next section we will introduce the structure of a single PerceiverIO block of our proposed architecture. We will then show how multiple blocks can be chained together to integrate information across all levels of the feature hierarchy. Finally, we will show how the overall architecture can be extended to allow the use of features from various different input modalities.

\subsubsection{Attention Mechanism}

Assume we have an input sequence, $I\in\mathcal{R}^{n \times d}$, with sequence length $n$ and embedding dimension $d$. As is standard~\cite{vaswani2017attention}, we define the attention mechanism as:
\begin{equation}
Attention(Q,K,V) = \phi (\frac{QK^T}{\sqrt{d}})V
\end{equation}
where $Q$, $K$ and $V$ are the query, key and value matrices respectively and $\phi()$ is the softmax operation. Normalization by the embedding dimensionality of the input features, $d$, is needed for training stability. In standard self-attention, the matrices $Q$, $K$ and $V$ are all derived from the input sequence $I$ via linear transformations of the sequence elements. Hence, $QK^T\in\mathcal{R}^{n \times n}$, which gives the self-attention mechanism its quadratic complexity with respect to sequence length. We will next show how a more efficient cross-attention mechanism can be implemented, allowing long-sequences to be processed.


\paragraph{Cross-Attention Block} Given a learned latent state matrix $z \in \mathcal{R}^{m \times k}$ and input sequence $x \in \mathcal{R}^{n \times d}$, we use the standard attention mechanism to perform cross-attention. A query matrix $Q\in \mathcal{R}^{m \times k}$ can be derived by linear transformation of the features in $z$. Similarly, a key matrix $K\in \mathcal{R}^{n \times k}$ can be derived from the input $x$. Now $QK^T\in\mathcal{R}^{m \times n}$ where $m << n$, making this operation much more efficient than standard self-attention.
The value matrix $V\in \mathcal{R}^{n \times k}$ is also derived from the input. The output matrix $\hat{z} \in  \mathcal{R}^{m \times k}$ from $\phi(QK^T)V$ is the updated latent state matrix containing a summary of information derived from the input.
We define a cross-attention block with a single attention head, as follows: 


\begin{lstlisting}[language=Python, caption=Cross-Attention Block.,numbers=none]
def CrossAttention(x,z):
    z = z + Attention(Norm(z).Wq,Norm(x).Wk,Norm(x).Wv) 
    z = Norm(z) 
    z = z + MLP(z) 
    return z
\end{lstlisting}

Where projection matrices $Wq\in \mathcal{R}^{k\times k}, Wk\in \mathcal{R}^{d\times d}, Wv \in \mathcal{R}^{d\times d}$, are used to linearly transform the elements of the latent state and input matrixes respectively. The query matrix is derived from the latent state matrix, $z$, while the key and value matrixes are derived from the input, $x$. A standard MLP layer consisting of two linear layers, each followed by dropout and a gelu activation is used~\cite{vaswani2017attention}. For explanation purposes we show here the case of a single attention head, however multiple attention heads can be used in practice.

The output from the cross-attention block is an updated latent state matrix $\hat{z} \in \mathcal{R}^{m \times k}$, which we interpret as containing a summary of the input. Note that the sequence length of the latent state is much shorter than the length of the input sequence i.e. $m << n$. The initial value of the latent state matrix passed to the first block is learned during training. The latent state matrix can now be passed to deeper self-attention layers for further processing.

Note - the dimensionality of the latent state embeddings, $k$, need not match the dimensionality of the input sequence embeddings, $d$. Reducing the dimensionality of the latent state embeddings significantly reduces the overall number of model parameters by reducing the size of all full-connected layers e.g., in the MLP and the projection matrixes using in the attention mechanism. In a departure from many other Transformer-based architectures, we use a latent state dimension that is significantly smaller than the input dimension. This ensures that our network can be trained from scratch in the small-data regime, commonly encountered with malware datasets (see Section~\ref{sec:experiments} for hyper-parameter details).


\paragraph{Self-Attention Block} The self-attention block is defined in a similar way to the cross-attention block. The attention operation itself does not change. Attention is performed between the latent state and itself. Given latent state matrix $z \in \mathcal{R}^{m \times k}$ the self-attention block is defined as:


\begin{lstlisting}[language=Python, caption=Self-Attention block.,numbers=none]
def SelfAttention(z):
    z = z + Attention(Norm(z).Wq,Norm(z).Wk,Norm(z).Wv)
    z = Norm(z)
    z = z + MLP(z)
    return z
\end{lstlisting}

Note that query, key and value matrixes are all derived from the latent state $z$, which attends with itself. As both the length of $z$ and its dimensionality are quite small, this operation is much more efficient than full self-attention applied directly to the input sequence. In effect, self-attention is applied to a summary of information from the input contained in $z$.

\paragraph{PerceiverIO Block} The basic unit of our proposed network is the PerceiverIO Block. This block consists of a cross-attention block followed by one or more self-attention blocks. The cross-attention block reads from the input opcode sequence to update the latent state. The latent state then contains a summary of the salient information extracted from the input sequence. The updated latent state is then passed to one of more self-attention blocks for further processing. Given latent state $z$ and input sequence $x$, we define the complete PerceiverIO block as follows:
%

\begin{lstlisting}[language=Python, caption=PerceiverIO Block.,escapeinside={(*}{*)}, numbers=none]
def PerceiverIOBlock(x,z):
    z = z + CrossAttention(x,z)
    for j = 1:J
        z = z + SelfAttention(*$_j$*)(z)
    return z
\end{lstlisting}

where $J$ is the number of self-attention blocks. We leave the decision about whether PerceiverIO blocks have shared parameters as an implementation detail. 

We note here that unlike many Transformer models we do not use positional embeddings. This is to allow for shuffling of the order of code blocks within an executable, which would have no effect on execution behaviour. Use of positional embeddings may still be useful in certain circumstances, such as encoding program control-flow-graph structure, however we leave this as an avenue for future work.

\subsection{Hierarchical Feature Combination}

Given the opcode-level features $C$, method-level features $M$, and the global feature, $g$, a chain of PerceiverIO blocks is used to combine the information to reach an overall malware classification decision. Separate PerceiverIO blocks are used for each feature level. 

The initial value of the latent state matrix is learned during training. To perform inference, the initial latent state matrix is passed to the first PerceiverIO block. Each PerceiverIO block in the chain then reads from its respective input feature, updates the latent state matrix using cross-attention, and performs processing using self-attention. It then passes the updated latent state matrix to the next PerceiverIO block in the chain. At the end of the chain, the latent state matrix now holds information derived from all the input features. It can now be used to make a malware classification decision (see Section~\ref{sec:classification}).

\subsection{Extension to Multiple Modalities}
\label{section:multimodal}

Our proposed architecture can be extended in a straightforward manner to use features from one or more additional modalities, e.g., requested app permissions, API calls, intents etc. Using additional modalities helps to increase the robustness of the classifier by providing independent evidence from multiple sources. Extension of the network to use requested permissions is shown in Fig.~\ref{fig:network_arch}.

For each additional modality an additional PerceiverIO block is added to the chain of PerceiverIO blocks. As before, each PerceiverIO block attends to its input modality, updates the latent state matrix, then passes it to the next block in the chain.

We assume that the features from each additional modality can be represented as a matrix $P\in\mathcal{R}^{\hat{n} \times \hat{d}}$, where $\hat{n}$ is the sequence length and $\hat{d}$ is the feature dimension. For example, the list of permissions requested by an app can be represented as a binary vector $\hat{P}\in\mathcal{R}^{1 \times \hat{p}}$, where $\hat{p}$ is the total number possible permissions. Modalities with both ordered and un-ordered features can be accommodated by choosing to use positional embeddings as required~\cite{vaswani2017attention}. Note that for each modality, the linear transformations generating the key and value matrices in its PerceiverIO block must be adapted to match the input dimensionality of the modality. No other changes to the basic architecture are required.

\subsection{Classification}
\label{sec:classification}

The latent state matrix output by the final PerceiverIO block contains information collected from all input modalities. We perform malware classification by applying Sequence Pooling~\cite{hassani2021escaping} to the latent state matrix output by the final PerceiverIO block i.e., Latent Pooling. Sequence pooling transforms the latent state matrix, $z \in \mathcal{R}^{m \times k}$, of sequence length, $m$, and embedding dimension, $k$, to a single vector, $\dot{z} \in \mathcal{R}^{1 \times k}$. This vector can then be used to perform classification. Sequence pooling works in a similar way to the attention mechanism itself. It allows the network to select the most relevant information from the latent state matrix to perform the classification task.

Each element of $z$ is first passed through a full-connected (i.e., linear) layer with output size $1$ to produce a vector $w \in \mathcal{R}^{m \times 1}$. The softmax operation is then applied to $w$ to normalise this vector. The normalized vector $w$ is then used to compute a weighted sum of the elements in $z$ i.e., $\hat{z} = w^T z$. Finally, the vector, $\dot{z} \in \mathcal{R}^{1 \times k}$, is passed to a linear classifier. Binary-cross-entropy loss is used when training the network to perform malware classification as this is a two-class problem with benign and malware classes.


\section{Experiments}
\label{sec:experiments}

In this section we investigate the performance of our proposed network. 
For all experiments involving the Malceiver the following hyper-parameters were used: Embedding Layer: $8$ dimensions. Number of convolutional filters can be varied between: $16, 32, 64, 128$. The default configuration with $64$ convolutional filters is used unless otherwise specified. Each convolutional filter was of length $8$ and width $8$. The PerceiverIO hidden state dimension was $8$. The latent state matrix was of length $64$ and dimensionality $8$. Weight sharing was not used between PerceiverIO blocks. Dropout of $0.1$ was used in MLP blocks. Self-attention blocks used 2 attention heads. Cross-attention blocks used 1 attention head. Each PerceiverIO block consisted of a single cross-attention block and a single self-attention block. These hyper-parameters, particularly the PerceiverIO hidden state dimension, ensure that the total number of learn-able parameters is kept small. Using the default hyper-parameters, with 64 convolutional filters, the complete Malceiver model has only 12,994 parameters.

At the start of training we use a warm-up procedure with a learning rate of 1e-5 for $25$ epochs. Cosine Annealing was then used during the main part of training with a period of $25$ epochs, a minimum learning rate of 1e-6, and a maximum learning rate of 1e-2. The batch size was $24$. The network was trained for $125$ epochs. All opcode sequences where of maximum length $128,000$ opcodes. Binary cross-entropy loss and Adam optimiser~\cite{kingma2014adam} were used.

We compare our system's performance with a baseline malconv malware detector~\cite{mclaughlin2017deep} trained on the same dataset and with the same split between training and testing data. The hyper-parameters of the baseline system were as follows: Embedding Layer: $8$ dimensions, Number of convolutional filters $64$ each filter was of length $8$ and width $8$, Max-Pooling layer used, followed by a single linear layer with a $1$ dimensional output. Binary cross entropy loss was used. Adam optimiser with batch size $24$, and learning rate of 1e-3 for $125$ epochs. The hyper-parameters used are the same as~\cite{mclaughlin2017deep}.

The dataset used for experiments was provided by McAfee Labs (Intel Security) and consists of malware from the vendor's internal dataset. This dataset contains roughly 10,000 malware and a further 10,000 clean applications collected from the Google Play Store. The datasets were cleaned to ensure no duplicate programs could contaminate the test/training splits. The dataset was split into 5 non-overlapping folds and cross validation was performed during evaluation. For each experiment we report the average performance across the 5 cross-validation folds in terms of f1-score. During training and testing, the length of all opcode sequences was truncated to 128,000 opcodes due to GPU memory limitations. This truncation only affects a very small number of programs, so has no significant effects on our results. 
Our networks were implemented in Pytorch~\cite{NEURIPS2019_9015}. We used distributed training over $3\times$Nvidia RTX6000 GPUs.

\subsection{Ablation}

We investigate the importance of the different hierarchical features derived from the opcode sequence by performing an ablation experiment. Three versions of our proposed network were trained and tested with three different feature combinations: opcode-features, opcode-features + method-features and opcode-features + method-features + global feature. In all cases, the number of convolutional filters was fixed at $64$. For each of the three feature combinations, the network was trained and tested using the same dataset splits to ensure a fair comparison. Results are reported in terms of average f1-score from 5-fold cross-validation in Table~\ref{table:ablation}.

\begin{table}[h]
\centering
\renewcommand{\arraystretch}{1.2}
\begin{tabular}{l|llll}
Features Combination     & Acc.   & Prec.  & Recall & F1-score \\ \hline
opcodes                  & 0.9328 & 0.9337 & 0.9304 & 0.9321 \\
opcodes + methods          & 0.9360  & 0.9389 & 0.9314 & 0.9351 \\
opcodes + methods + global & 0.9469 & 0.9489 & 0.9436 & 0.9462
\end{tabular}
\vspace{5mm}
\caption{Ablation experiment where different combinations of features, derived from the opcode sequence, are tested to study how this affects malware classification performance. Results are reported in terms of average f1-score from 5-fold cross-validation.\label{table:ablation}}
\end{table}

From the results in Table~\ref{table:ablation} we can see that adding additional feature types helps improve overall malware classification performance. We note that method features only give a small improvement over opcodes. There is a much larger improvement when all three feature types are used. It may be the case that much of the information contained in the method features has already been captured by the original opcode level features. This may be caused by the closeness of the window length of the convolutional filters to the average method length.
The global feature summarises long range information across the complete program length. Consequently, we see that the addition of this feature significantly improves performance compared with the purely local opcode features used alone.

\subsection{Hyper-parameters}

The perceiver architecture has several internal hyper-parameters that can be varied. Due to computational restrictions, we cannot investigate every hyper-parameter fully. For most hyper-parameters we try to follow guidelines established in the literature. We choose to investigate two hyper-parameters important to the PerceiverIO architecture: the length and the dimensionality of the latent state matrix $z \in \mathcal{R}{m\times k}$, where $m$ is the length and $k$ is the dimensionality. Changing the dimensionality of $z$ has a significant effect on the number of network parameters, as it affects all cross-attention, self-attention and MLP layers. Changing the length of the latent state affects the amount of 'memory' available to the network, while having only a minor effect of parameter count.

The default latent state matrix is of length $64$ and dimensionality $8$. We vary the length between $16$ and $128$, while keeping the latent state dimensionality fixed at the default value of $8$. And we vary the dimensionality between $2$ and $32$, while keeping the latent state length fixed at the default value of $64$. We also conduct and experiment where we vary both parameters together from $2$ to $64$. All other network parameters were unchanged. We report results in Table~\ref{table:per_dimensionaltiy}, Table~\ref{table:per_length} and Table~\ref{table:per_length_dim}.

\begin{table}[]
\centering
\renewcommand{\arraystretch}{1.2}
\begin{tabular}{l|llll}
Latent State Dimensionality & Acc.   & Prec.  & Recall & F1-score \\ \hline
2           & 0.5006 & 0.1971 & 0.3172 & 0.2393 \\
4           & 0.9396 & 0.9453 & 0.9319 & 0.9386 \\
8           & 0.9413 & 0.9462 & 0.9345 & 0.9403 \\
16          & 0.9408 & 0.9473 & 0.9324 & 0.9397 \\
32          & 0.9406 & 0.9439 & 0.9358 & 0.9398
\end{tabular}
\vspace{5mm}
\caption{Hyper-parameter experiment varying the dimensionality of Perceiver hidden state matrix. Results are reported in terms of the average accuracy, precision, recall and f1-score from 5-fold cross-validation.\label{table:per_dimensionaltiy}}
\end{table}

\begin{table}[]
\centering
\renewcommand{\arraystretch}{1.2}
\begin{tabular}{l|llll}
Latent State Length & Acc.   & Prec.  & Recall & F1-score \\ \hline
16          & 0.9423 & 0.9489 & 0.9338 & 0.9413 \\
32          & 0.9420 & 0.9447 & 0.9378 & 0.9412 \\
64          & 0.9408 & 0.9457 & 0.9342 & 0.9399 \\
128         & 0.9408 & 0.9437 & 0.9364 & 0.9400 
\end{tabular}
\vspace{5mm}
\caption{Hyper-parameter experiment varying length of Perceiver hidden state matrix. Results are reported in terms of the average accuracy, precision, recall and f1-score from 5-fold cross-validation.\label{table:per_length}}
\end{table}

\begin{table}[]
\centering
\renewcommand{\arraystretch}{1.2}
\begin{tabular}{ll|llll}
Latent State Length & Latent State Dimensionality & Acc.   & Prec.  & Recall & F1-score \\ \hline
2                   & 2                           & 0.5048 & -      & -      & -        \\
4                   & 4                           & 0.9411 & 0.9464 & 0.9339 & 0.9401   \\
8                   & 8                           & 0.9391 & 0.9449 & 0.9314 & 0.9381   \\
16                  & 16                          & 0.9384 & 0.9424 & 0.9328 & 0.9375   \\
32                  & 32                          & 0.9410 & 0.9483 & 0.9317 & 0.9399   \\
64                  & 64                          & 0.8077 & 0.6957 & 0.7500 & 0.7170  
\end{tabular}
\vspace{5mm}
\caption{Hyper-parameter experiment varying both length and dimensionality of Perceiver hidden state matrix together. Results are reported in terms of the average accuracy, precision, recall and f1-score from 5-fold cross-validation. (Note that when both values were two, training failed to converge.)\label{table:per_length_dim}}
\end{table}

From Table~\ref{table:per_dimensionaltiy} and Table~\ref{table:per_length} We can see that in the ranges tried, the network is robust to variation of these hyper-parameters. The only exception is when the latent state dimensionality is reduced significantly performance collapses. This is likely because the network cannot represent the necessary information to make an accurate decision. This result is confirmed in Table~\ref{table:per_length_dim}, where both parameters are varied together. Performance is generally stable. However for very small values of the parameters, training fails to converge. For larger values of the parameters performance significantly declines.

Note that in order to fully understand the significance of the hyper-parameters more extensive evaluation of different combinations would be required. It is likely that information can be passed through the network in different ways. It remains unclear whether latent state dimensionality or length is more important. In the rest of the experiments we use the default hyperparameter values. As we have shown here, small changes in these hyper parameters would not significantly affect results. The default hyper-parameter settings ensure the number of learned parameters is kept low, which should help prevent over-fitting given the relatively small malware datasets used.

\subsection{Multiple Modalities}


As mentioned in Section~\ref{section:multimodal}, our architecture can be extended to use evidence from multiple modalities for malware classification. This can be done by adding extra PerceiverIO blocks, one for each modality. In this experiment we use requested app permissions in addition to opcode-sequences. We note that our approach is flexible and can be extended to include any number of additional modalities available.

For each APK file in our dataset we extract a list of 138 requested permissions, based those used in~\cite{6531746}, which showed that permissions could give accurate malware classification performance for Android APKs. A network was trained to perform malware classification using the permission features in addition to the opcode-sequence features in all other experiments. The complete network architecture is shown in Fig~\ref{fig:network_arch}. 
In common with our previous experiment (see Section~\ref{sec:comp_sota}) we vary the number of convolutional filters in the CNN layer. This allows us to measure the utility of the combined permission and opcode features across a range of network sizes. For all experiments we report average F1-score from 5-fold cross validation. The same training and testing splits were used for all network variants.

We first performed a baseline experiment where only permission features were used. In this case only a single PerceiverIO-block, taking permissions as input, was used. Opcode features were not used. The malware classification f1-score of our system using permission features alone was 0.8391.

Next we performed multi-modal experiments using opcode and permission features. Results for the multi-modal system are shown in Table~\ref{table:multimodal}. We can see that using permission features in addition to opcodes improves the system's performance in all cases. This is despite the relatively low performance with permission features used alone. Permissions and opcodes provide complementary information, allowing errors to be corrected that would not be possible with either feature used alone.  The use of permission features gives a larger improvement in cases where the underlying network has a smaller number of convolutional filters. As the number of CNN filters is increased, the system's performance starts to saturate and the use of permission features gives less of a boost in f1-score compared to opcode-sequence features alone. This likely indicates we are approaching the limits of possible classification performance on this dataset. 

Overall this experiment demonstrates the ability of our architecture to make use of information from multiple modalities to improve malware classification performance. This has been shown over a range of network sizes.

\begin{table}[]
\centering
\begin{tabular}{r|llll|llll}
                  & \multicolumn{4}{c|}{Opcode Features} & \multicolumn{4}{c}{Opcode \&   Permission Features} \\ \hline
Num. Conv Filters & Acc.    & Prec.  & Recall & F1-score & Acc.       & Prec.      & Recall     & F1-score    \\ \hline
16                & 0.9324  & 0.9352 & 0.9278 & 0.9315   & 0.9423     & 0.9432     & 0.9401     & \textbf{0.9417}      \\
32                & 0.9393  & 0.9433 & 0.9335 & 0.9384   & 0.9471     & 0.9492     & 0.9438     & \textbf{0.9465}      \\
64                & 0.9470  & 0.9473 & 0.9456 & 0.9465   & 0.9531     & 0.9626     & 0.9420     & \textbf{0.9521}      \\
128               & 0.9533  & 0.9558 & 0.9497 & 0.9527   & 0.9568     & 0.9633     & 0.9490     & \textbf{0.9561}     
\end{tabular}
\vspace{5mm}
\caption{Comparing performance of our proposed system using opcode features only and using multi-modal features i.e., opcodes and requested permissions. The number of convolutional filters was also varied. Results are reported in terms of the average accuracy, precision, recall and f1-score from 5-fold cross-validation. The highest f1-scores are shown in bold. \label{table:multimodal}}
\end{table}

\subsection{Comparison with Literature}
\label{sec:comp_sota}

In this experiment we compare the performance of our proposed Malceiver architecture with the widely used Malconv opcode-sequence based malware classifier~\cite{mclaughlin2017deep}. We report performance for the Malceiver using both opcode-sequence features only, and using opcode-sequence features combined with requested permissions. To ensure a fair comparison all systems were trained and tested on the same dataset.

It is known from existing work on CNNs for opcode-sequence based malware classification that changing the number of convolutional filters can have an influence on classifier performance~\cite{mclaughlin2017deep}. Each CNN filter acts as an opcode pattern detector. Adding more filters - until just before the point of over-fitting - allows the network to detect a wider range of malware. The network's deeper layers then combine the evidence about which patterns were present to reach a classification decision. 

\begin{figure}[h!]
\centering
\includegraphics[width=0.8\textwidth,trim={2cm 2.5cm 2cm 2.5cm},clip]{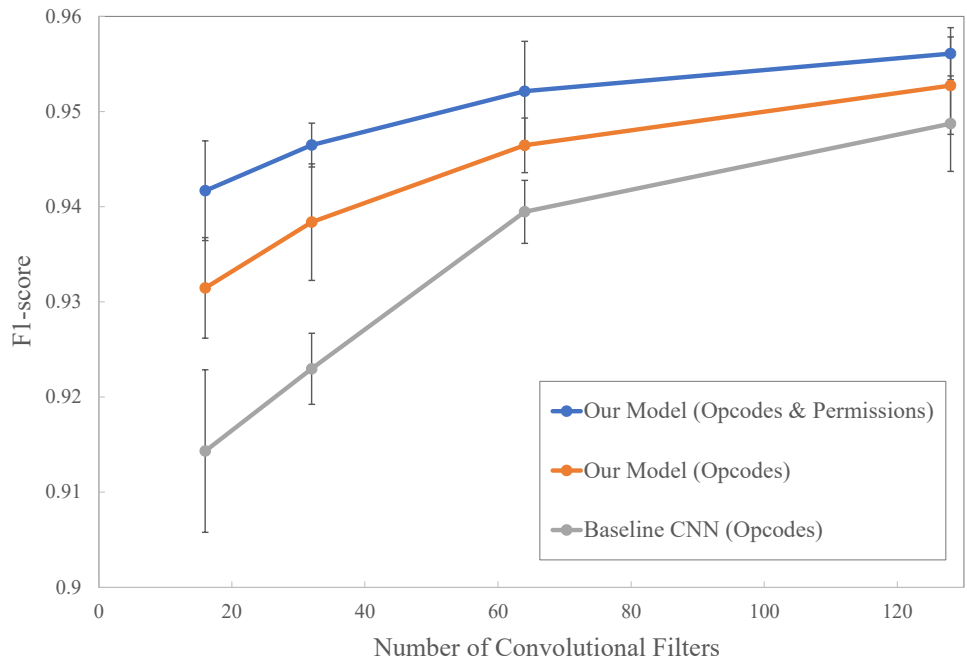}
\caption{Over a range of convolutional filters numbers (16, 32, 64, 128) we report the mean and standard deviation of the f1-score on the  validation-set for: Our Model (Malceiver) with opcodes, Our Model with opcodes and permission features, and the baseline CNN model from the literature~\cite{mclaughlin2017deep} using opcode features only.\label{fig:perf_comp}}
\end{figure}

For both the baseline CNN and Malceiver, the number of convolutional filters was varied between $16, 32, 64$ and $128$. This experiment shows the sensitivity of our proposed architecture to the size of the initial CNN network. For each network and for each number of convolutional filters we report the mean and standard deviation F1-score from 5-fold cross-validation. The same dataset and same training and testing splits were used for all networks. Results are reported in Figure~\ref{fig:perf_comp}.

Firstly, we see that increasing the number of convolutional filters helps improve the classification performance of all models. We also see that classification performance is asymptotic. It appears that increasing the number of convolutional filers beyond what was tried is unlikely to give substantial performance improvements.

Secondly, we can see that for any given number of convolutional filters the Malceiver outperforms the Baseline 1D CNN model. This is true with both opcode features only and with combined opcode and permission features. With opcode features only, both the Malceiver and baseline network have the same raw information available. However, the Malceiver appears to be better able to use this information for accurate decision making. This could due to its ability to combine information across the features hierarchy, such as combining local with global features.

Overall we demonstrate the potential of the Malceiver architecture. This architecture is able to consistently outperform the baseline across a range of network sizes, even with the same input features. We then show that by adding additional multi-modal features performance further improves. This demonstrates both the flexibility of our architecture and its potential for future performance gains with additional complementary modalities.

\section{Conclusion}

In this paper we have shown, for the first time, how transformer-based models can be efficiently applied to static-analysis malware detection for Android using full-length opcode-sequences. To do this we adapt the Perceiver/PerceiverIO architecture to the malware classification task. We have shown how our Malceiver model can be extended to use multiple modalities for improved performance. 
The Malceiver architecture is more flexible than previous opcode-sequence malware classification architectures, based on image-classification networks or 1D CNNs. The new architecture opens up the possibility of training larger models capable of making use of ever larger datasets incorporating as many additional modalities as available.



\bibliographystyle{unsrt}  
\bibliography{references}  

\end{document}